\documentclass[a4paper]{article}
\usepackage{geometry}
\usepackage{authblk}
\usepackage{hyperref}
\usepackage{graphicx}
\usepackage{caption}
\usepackage{natbib}
\usepackage{multicol}
\usepackage{multirow}
\usepackage{booktabs}
\usepackage{aas_macros}
\newcommand\MyHead[2]{%
    \multicolumn{1}{l}{\parbox{#1}{\centering #2}}
}
\newcommand{\myimageTwo}[2]{\begin{center}\includegraphics[angle=0, width=0.45\textwidth]{images/#1}
        \includegraphics[angle=0, width=0.45\textwidth]{images/#2}\end{center}}

\title{Isochrone fitting in the Gaia era. II. Distances, ages and masses from UniDAM using Gaia DR2 data.}
\author[1, 2]{A. Mints}
\affil[1]{Max Planck Institute for Solar System Research, Justus-von-Liebig-Weg 3, 37077 Göttingen, Germany}
\affil[2]{Stellar Astrophysics Centre, Department of Physics and Astronomy, Aarhus University, Ny Munkegade 120, DK-8000 Aarhus C, Denmark}
\begin{document}

\maketitle

\begin{abstract}
We present estimates of distances, ages and masses for over 3.5 million stars. These estimates are derived from 
the combination of spectrophotometric data and Gaia DR2 parallaxes.
For that, we used the previously published Unified tool to estimate Distances, Ages, and Masses (UniDAM).
\end{abstract}

\section{Introduction}
In \cite{2018arXiv180406578M} we presented an update of the Unified tool to estimate Distances, Ages, and Masses (UniDAM\footnote{UniDAM source code is available at \url{https://github.com/minzastro/unidam}.}) \citep{2017A&A...604A.108M}, which allowed for the use of parallax data in isochrone fitting.
Recent release of Gaia DR2 \citep{2016A&A...595A...1G,2018arXiv180409365G} allowed us to produce 
a new catalogue of distance, ages and masses for over 3,5 million stars. We were also able
to test the predictions for distance and age uncertainties described in \cite{2018arXiv180406578M}.

As compared to \cite{2018arXiv180406578M} our sample was extended with the second data release from GALAH \citep{2018arXiv180406041B}, that contains parameters for nearly 350,000 stars.
Our sample is summarized in the Table \ref{tbl:overlap} below.

\begin{table*}[t]
    \centering
            \begin{tabular}{lrrr} \toprule
    Survey              & \MyHead{2cm}{Input catalog size}     & 
    \MyHead{2cm}{Estimates without parallaxes} & 
    \MyHead{2cm}{Estimates with Gaia DR2 parallaxes}  \\ \midrule
    APOGEE (DR14)       & 157,322   & 149,599             & 139,253   \\
    Gaia-ESO (DR2)      & 6,376     & 5,964               & 5,140     \\
    GALAH (DR1)         & 342,682   & 335,783             & 321,836   \\
    GCS                 & 13,565    & 12,987              & 8,079     \\
    LAMOST (DR3)        & 3,036,870 & 2,957,410           & 2,607,581 \\
    LAMOST\_CANNON*     & 444,784   & 436,617             & 406,870   \\
    LAMOST GAC (DR2)*   & 366,173   & 354,409             & 321,194   \\
    LAMOST GAC VB (DR2)*& 1,063,950 & 1,033,650           & 927,972   \\
    RAVE (DR5)          & 491,349   & 451,587             & 347,741   \\
    RAVE\_on*           & 440,913   & 427,009             & 362,405   \\
    SEGUE               & 235,595   & 202,162             & 108,471   \\
    TESS-HERMES (DR1)   & 15,872    & 15,657              & 14,669    \\ \midrule
    Total               & 4,249,195 & 4,106,571           & 3,567,434 \\ \bottomrule
            \end{tabular}
    \caption{Total number of sources and Gaia DR2 overlap for different surveys. *- LAMOST GAC, LAMOST-Cannon and RAVE-on were processed but not included into the total, as they contain the same stars as LAMOST DR3 and RAVE DR5.}\label{tbl:overlap}
\end{table*}

\section{Results and discussion}
We publish results of UniDAM with Gaia DR2 parallaxes on \url{http://www2.mps.mpg.de/homes/mints/unidam.html}. 

In Figure \ref{fig:results} we show how median uncertainties of distance modulus and log(age) derived with the use of Gaia DR2 data compare with results without parallaxes and with predictions for Gaia end-of-mission (EoM) parallax quality from \cite{2018arXiv180406578M}. In most cases, distance modulus uncertainties obtained with Gaia DR2 data are larger than those expected for Gaia EoM. This is as expected, given that Gaia DR2 parallaxes do not yet reach the EoM precision.

As for log(age) uncertainties, they are in most cases comparable with predictions. There is a trend for log(age) uncertainties to be smaller than expected for nearby stars and larger than expected for most distant ones (see, for example, figure \ref{fig:results}h). Further analysis is needed to explain this behaviour.

For the majority of the surveys we find a consistent solutions for 85 to 95\% of the stars. Only for RAVE we find consistent solutions for 75\% of the stars.

\section*{Acknowledgements}
The research leading to the presented results has received funding from the European Research Council under the European Community's Seventh Framework Programme (FP7/2007- 2013)/ERC grant agreement (No 338251, StellarAges). 

This work has made use of data from the European Space Agency (ESA) mission
\textit{Gaia} (\url{https://www.cosmos.esa.int/gaia}), processed by the {\textit Gaia}
Data Processing and Analysis Consortium (DPAC,
\url{https://www.cosmos.esa.int/web/gaia/dpac/consortium}). Funding for the DPAC
has been provided by national institutions, in particular the institutions
participating in the {\it Gaia} Multilateral Agreement.

\begin{figure}
    \myimageTwo{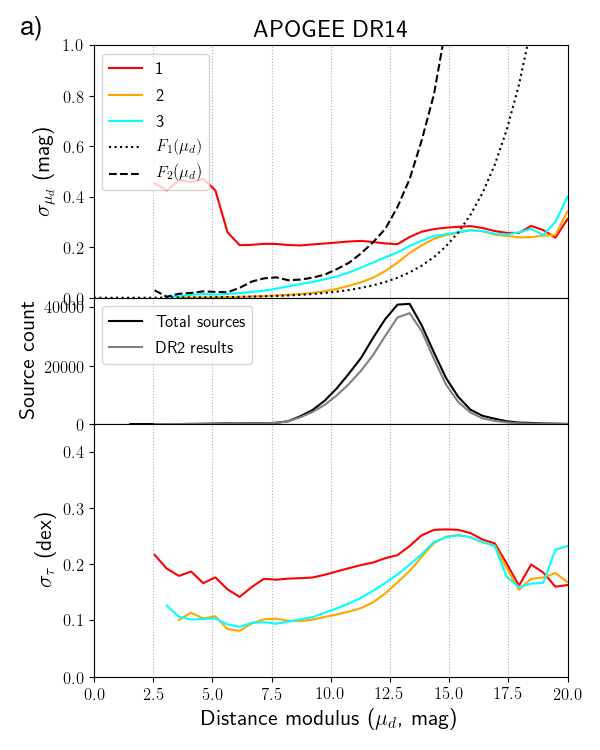}{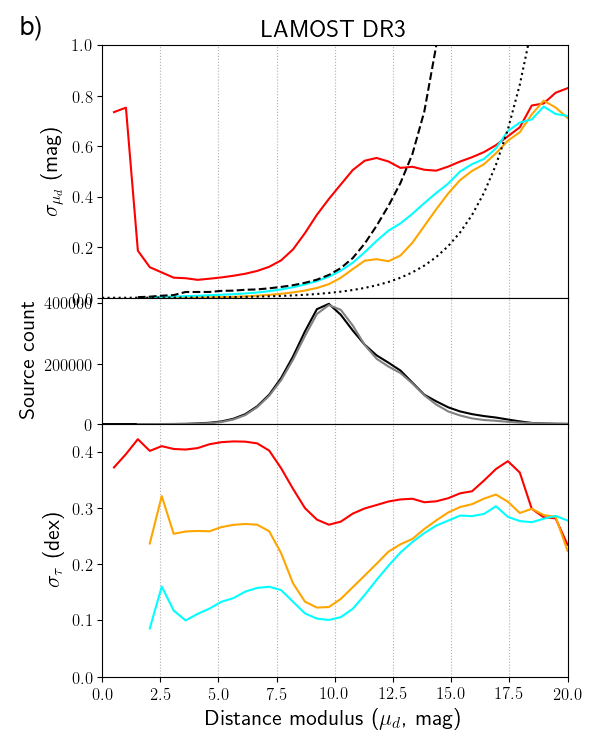}
    \caption{\textit{Top}: distance modulus uncertainties $\sigma_{\mu_d}$ as functions of distance modulus. Legend: 1 -- results without the use of parallaxes. 2 -- predictions for Gaia end-of-mission parallax quality from \cite{2018arXiv180406578M}. 3 -- results with Gaia DR2 parallax values. The dotted black line shows $F_1(\mu_d)$, the dashed black line shows $F_2(\mu_d)$ (as described in Section 4 of \cite{2018arXiv180406578M}). 
        \textit{Middle:} total number of sources (black) and number of sources with Gaia DR2 results (grey) as a function of distance modulus. \textit{Bottom:} log(age) uncertainties $\sigma_{\tau}$ as functions of distance modulus. }\label{fig:results}
\end{figure}
\begin{figure}\ContinuedFloat
    \myimageTwo{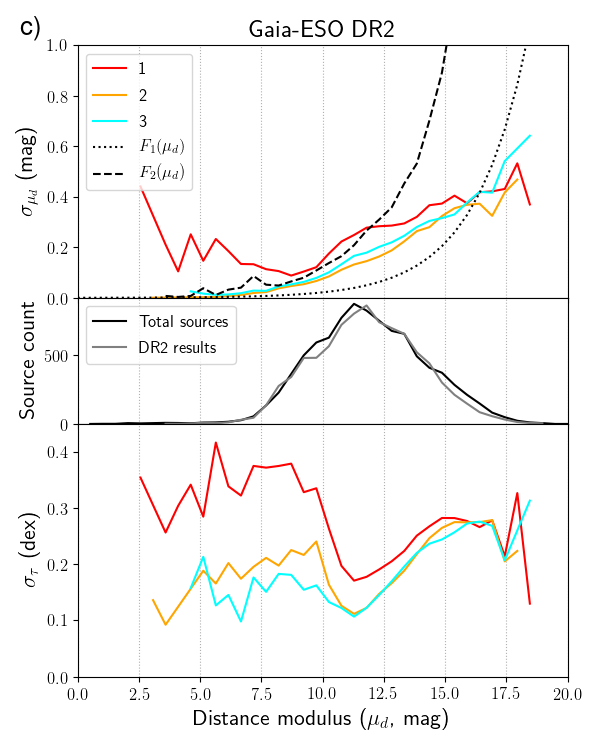}{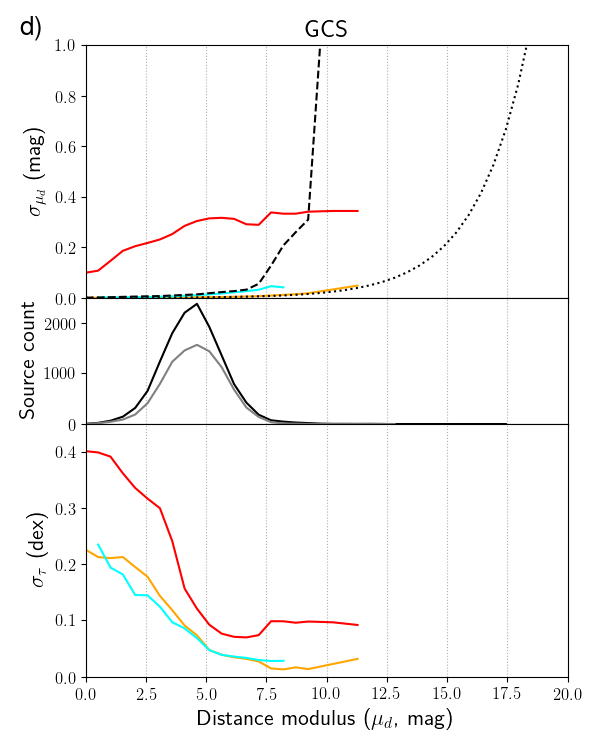}\par
    \myimageTwo{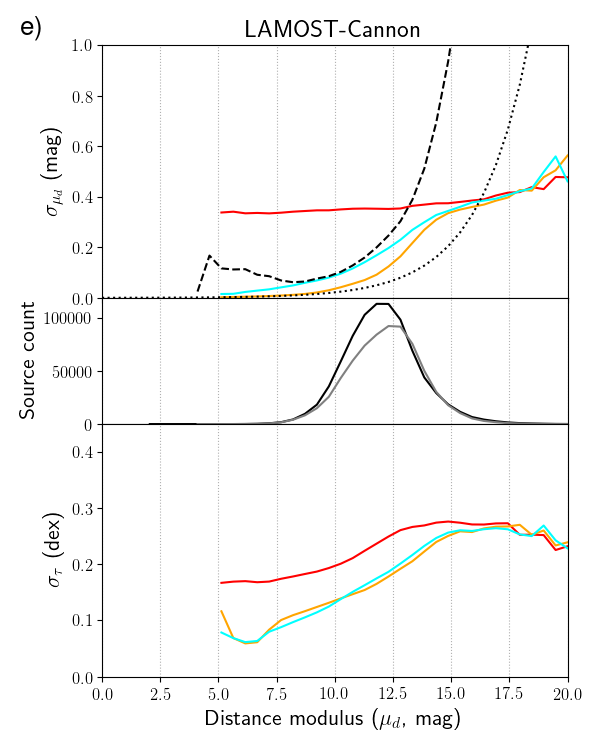}{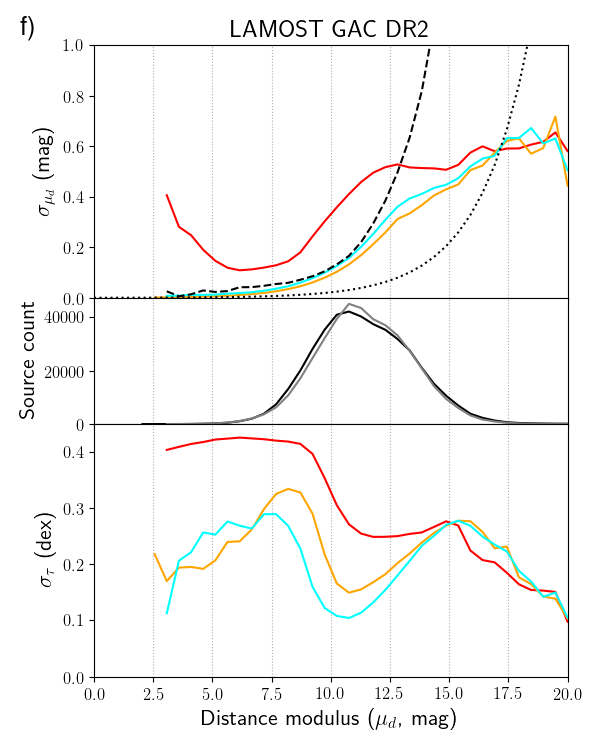}
    \caption{continued}\label{fig:unc3}
\end{figure}
\begin{figure}\ContinuedFloat
    \myimageTwo{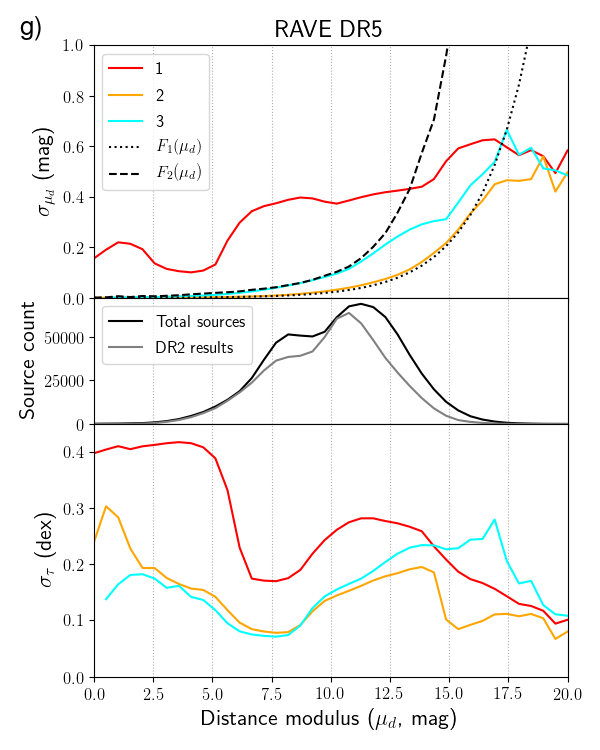}{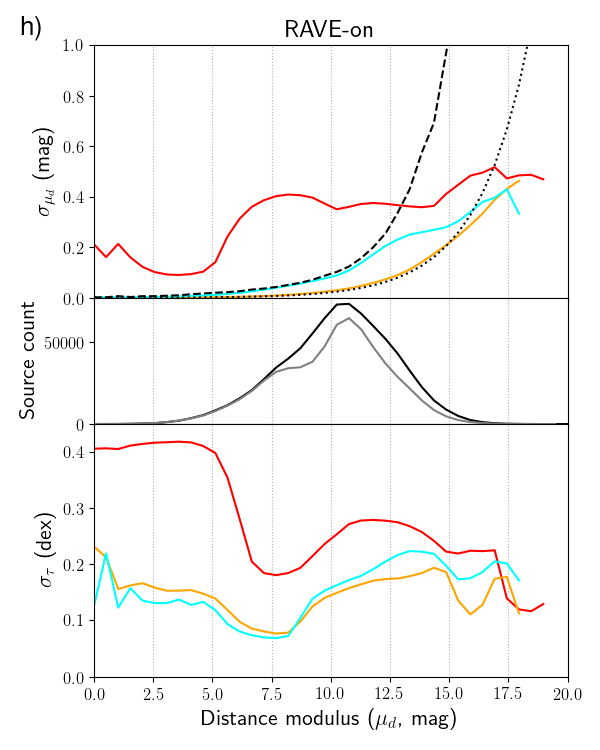}\par
    \myimageTwo{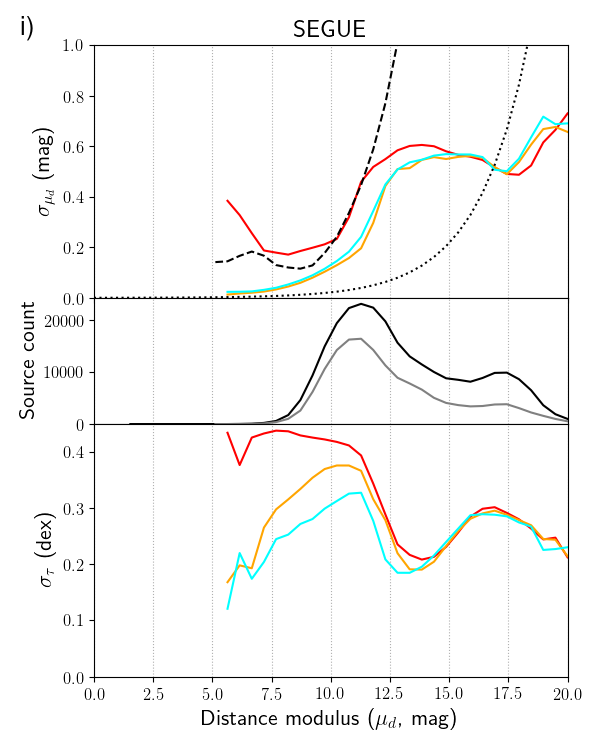}{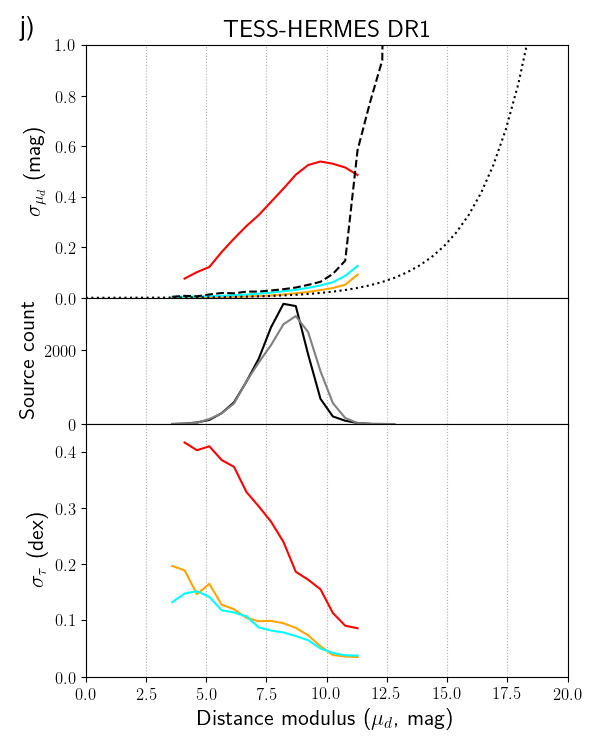}
    \caption{continued}\label{fig:unc4}
\end{figure}
\bibliographystyle{aa}
\bibliography{bibliography}

\end{document}